
\magnification=\magstep1
\hsize=159.2 true mm
\vsize=236 true mm
\baselineskip=19.1pt
\overfullrule=0pt
\font\heading=cmbx12 scaled \magstep1
\font\ninerm=cmr9
\font\ninebf=cmbx9
\font\ninett=cmtt9
\font\elevenbf=cmbx9 scaled \magstep1
\font\twelvebf=cmbx12
\newcount\figno
\def\no{\global\advance\figno by 1 \the\figno}
\global\figno=0
\newcount\equationno
\def\eqn{\global\advance\equationno by 1 \eqno(\the\equationno)}
\global\equationno=0

\leftline{\heading Anisotropic Finite-Size Scaling Analysis}
\medskip
\leftline{\heading of a Two-Dimensional Driven Diffusive System}
\bigbreak
\leftline{\elevenbf Jian-Sheng Wang\footnote{$^*$}{\ninerm Computational
Science, Blk S16, National University of Singapore, Lower Kent Ridge
Road, Singapore 0511.  E-mail: {\ninett cscwjs@leonis.nus.sg}}}
\bigskip
{\ninerm\narrower

\hbox to \hsize {\hskip\leftskip\hrulefill\hskip\rightskip}

\noindent The two-dimensional uniformly driven diffusive system is
simulated with a multi-spin coding technique.  The nonequilibrium
phase transition is analyzed with anisotropic finite-size scaling,
both at the critical point and off the critical point.  The
field-theoretical values of critical exponents fit the data well.

\hbox to \hsize {\hskip\leftskip\hrulefill\hskip\rightskip}

\noindent{\ninebf Key Words:}  Driven diffusive systems; anisotropic
finite-size scaling; nonequilibrium phase transitions; computer
simulations.\par}

\bigskip
\medskip
{\noindent\twelvebf 1. Introduction\par}
\nobreak\smallskip\nobreak
\noindent Driven diffusive systems are a class of models which exhibit
nonequilibrium phase transitions.$^{(1-4)}$ The models are defined by
some local rules similar to Kawasaki dynamics.  The steady states of
the dynamical evolution have been studied extensively during past
decade.  A central issue is whether the concept of universality of
critical phenomena can be applied in nonequilibrium cases.

The standard driven diffusive model of half-filled charged lattice
gas was proposed as a model for ionic solution in an electric
field.$^{(1,2)}$  A continuum version, based on symmetries and
conservation laws, was solved in a field-theoretic
framework.$^{(5,6)}$  It is quite remarkable that exact critical
exponents are obtained for dimensions from two to five.  In
particular, the set of critical exponents in two dimensions is
$$\beta = { 1\over 2}, \qquad
\gamma = 1, \qquad
\nu_\parallel = { 3\over 2}, \qquad
\nu_\perp = { 1\over 2}. \eqn $$
These exponents have similar meanings as in equilibrium second-order
phase transitions.  They are thought to be universal within a class.
No independent methods have been applied to the problem to check out
the validity of the claim except computer simulations.$^{(7,8)}$
Anisotropic finite-size scaling studies by Leung$^{(9,10)}$ appear to
have settled a dispute between theory and early computer simulations.
But a recent work$^{(8)}$ casts doubt on Leung's conclusion.  We
present here high quality data to support Leung's results.

\bigbreak
{\noindent\twelvebf 2. Model and Simulation\par}
\nobreak\smallskip\nobreak
The system consists of a square lattice of $L \times M$ sites.  The
driven field is in $x$ direction.  A configuration is specified by a
set of Ising spins, $\sigma_{x,y} = \pm 1$, with zero total
magnetization.  The state evolves according to the following
prescription.  A bond is chosen with equal probability in orientations
and in locations.  If the bond is parallel to the driven field and the
adjoining spins are distinct, the spins are changed to $\sigma_{x,y} =
-1$ and $\sigma_{x+1 \bmod L, y} = +1$.  This corresponds to an
infinitely strong driven field.  If the bond is perpendicular to the
field, we swap the spin values with a probability
$\min\bigl\{1, \exp(-\delta E/k_BT)\bigr\}$, where $\delta E$ is the
energy increment due to the change, assuming the usual nearest
neighbor ferromagnetic interaction with coupling constant $J$ and
periodic boundary conditions.  One Monte Carlo step is defined as
$L\times M$ such basic steps.

Simulation near critical point is not easy because of critical slowing
down.  The situation for this model is more severe by the conservative
nature of the dynamics, leading to relaxation time $\tau \propto M^4$.
Thus, an efficient implementation is crucial to obtain good
statistics.  With a multi-spin coding method,$^{(11)}$ thirty two
systems are simulated simultaneously.  This gives us at least a factor
of 10 speed up.  A slight penalty of the multi-spin coding algorithm
is that the temperature can not be set exactly to the desired value.
But it can be well under control with a large random bit table.  To
achieve an accuracy of five significant figures, we took a random bit
table of $2^{18}$ entries.

The program runs at 0.3 $\mu$sec per spin exchange on an SGI Iris
Indigo Workstation.  Computations were done on a cluster of
over fifty workstations (SGI Iris Indigo, HP 9000/700 model 712, and
DEC 5000) in two months.  The lengths of the runs are $10^7$ to $10^8$
Monte Carlo steps.  These are orders of magnitude longer than previous
studies.  Measurements are performed at an interval of 10 Monte Carlo
steps.

We use the order parameter introduced by Binder and Wang$^{(12)}$ and
modified by Leung.$^{(9)}$  Let's define
$$ \phi = {1\over 2 L} \sin\left({\pi\over M}\right)\;
\Bigl| \sum_{x=0}^{L-1} \sum_{y=0}^{M-1}
\sigma_{x,y} e^{i2\pi y/M} \Bigr|. \eqn $$
The normalization is such that $\phi = 1$ for a strip geometry (the
configuration in the limit $T \to 0$).  The following quantities are
calculated, (1) the order parameter $\Psi = \langle \phi \rangle$,
(2) the ``susceptibility,'' or fluctuation of the order parameter,
$$ \chi = {2 L \over \sin{\pi/M}}\; \Bigl[ \langle \phi^2 \rangle
 -  {\langle \phi \rangle}^2 \Bigr],\eqn$$
and the susceptibility above the critical temperature,
$$ \chi' = {2 L\over \sin{\pi/M}}\; \langle \phi^2 \rangle, \eqn$$
and (3) the fourth-order cumulant,
$$g=2-{\langle \phi^4 \rangle \over {\langle\phi^2 \rangle}^2}.\eqn$$
Note that $g$ goes from 0 to 1 as temperature $T$ goes from $\infty$
to 0.  We'll measure temperature in units of the two-dimensional Ising
critical temperature ($2.269 J/k_B$).

\bigbreak
{\noindent\twelvebf 3. Determination of the Critical Temperature\par}
\nobreak\smallskip\nobreak
Estimating the critical exponents depends crucially on an accurate
value of critical temperature.  In the literature different authors
have given incompatible values, $T_c = 1.355 \pm 0.003$ by Vall\'es
and Marro,$^{(7)}$ $T_c = 1.30\pm 0.01$ by Achahbar and
Marro,$^{(8)}$ and Leung's result$^{(9)}$ of $T_c = 1.418\pm 0.005$.
The discrepancies are the manifestation of the difficulties of
simulating the system and of interpreting the complicated finite-size
data.

We determine $T_c$ using two independent methods.  First, we look at
the peak of the susceptibility $\chi$.  The method is not
particularly accurate, but we get a general picture of the
finite-size critical temperature $T_c(L,M)$.  For very elongated
shapes, $T_c(L,M)$ decreases towards zero.  For a fixed value $M$ or
$L$, $T_c(L,M)$ reaches its maximum at about $S = L^{1/3}/M \approx
0.2$.  The systems with ratio $S \approx 0.2$ notice the finiteness of
the sizes in two directions at roughly the same temperature.
$T_c(L,M)$ increases as system size increases with fixed $S$.
Figure~\no\ is a plot of the peak locations $T_c(L,M)$ versus
$L^{-2/3}$, for some fixed values $S$ as well as for the square
geometry $L = M$.  According to the usual finite-size scaling
assumption, we should expect
$$ T_c(L, M) = T_c(\infty,\infty) + F(S) L^{-1/\nu_\parallel}. \eqn $$
Clearly, the data do not follow this equation very well, presumably
due to corrections to scaling.  But linear extrapolations should give
us lower bounds for the infinite system transition temperature $T_c$.
We quote the following result,
$$ T_c = 1.410 \pm 0.006. \eqn $$
It agrees with Leung's result$^{(9)}$ within errors.  For the square
systems, the convergence to the critical temperature is extremely
slow.  This may explain why the previous calculations on square
systems all gave a lower $T_c$.  If anisotropy is a dominant feature,
we should not expect peak locations to scale simply as $L^{-1/\nu}$
for square systems.

The peaks were not located with great precise because the
simulations were carried out at discrete points spaced at $\Delta T =
0.01$.  The second method exploits the scaling properties of the
fourth-order cumulant.  From finite-size scaling theory, we
have
$$g(T,L,M)=\tilde g\bigl(L^{1/\nu_\parallel}(T-T_c)/T_c, S\bigr).\eqn$$
If scaling were exactly obeyed, different curves (with fixed $S$)
should intersect at exactly the same value $T_c$.  Therefore, there is
no need to extrapolate to infinite size.  In practice there are
unknown corrections to scaling.  A better way is to consider the
overall scaling, Eq.~\the\equationno.  The value of $T_c$ for each $S$
can be determined more precisely.  However, there are weak size and
$S$ dependence.  Nevertheless, we found that the values all fall into the
interval $1.395$ to $1.410$.  Figure~\no\ is one of the scaling plots
with $S = 2^{-8/3}$.

Even with much greater computational effort, the accuracy of $T_c$ has
not been improved.  Each set of data or each method may give more
accurate value, but different sets of data or methods give slightly
different values.

\bigbreak
{\noindent\twelvebf 4. Anisotropic Finite-Size Scaling at Critical Point}
\nobreak\smallskip\nobreak
The finite-size scaling theory of isotropic systems can be generalized
to anisotropic systems.$^{(12)}$ The idea is that the anisotropic
system has the same scaling form as the isotropic one if we fix the
ratio $S = L^{\nu_\perp / \nu_\parallel} / M$.  In anisotropic
finite-size scaling, $S$ enters as an independent variable, in
addition to the usual scaling variable $L^{1/\nu_\parallel}
(T-T_c)/T_c$.  particularly, at $T=T_c$, scaling functions depend on
$S$.

In applying the finite-size scaling theory, we could simulate a very
large system and study the size effect of smaller subsystems.  This
may seem computationally effective.  But there are two problems
associated with it: we have the annoying finite-size effect of the
very large system when the subsystem sizes are comparable to it; and
we may not be able to equilibrate the very large system very well.
So, we adopt the more conventional finite-size scaling
analysis---working on fully finite size of dimension $L \times M$.

The exponent ratio $\nu_\parallel/\nu_\perp$ is one of the most
important number in an anisotropic finite-size analysis.  Theoretical
result is often assumed.$^{(9,13)}$ We have attempted to determine it
from our data.  The strip geometry with $S \to 0$ or $S \to
\infty$ and periodic boundary conditions has a simpler scaling
behavior at the critical temperature,$^{(12)}$
$$\chi(T_c)  \propto M^{\gamma / \nu_\perp}, \qquad
\Psi(T_c) \propto M^{\gamma/ 2\nu_\perp - 1/2} L^{-1/2}, \qquad L \gg
M^{\nu_\parallel/\nu_\perp}, \eqn$$
$$\chi(T_c)  \propto L^{\gamma/\nu_\parallel}, \qquad
\Psi(T_c) \propto L^{\gamma/2\nu_\parallel - 1/2} M^{-1/2},
\qquad M \gg L^{\nu_\perp/\nu_\parallel}. \eqn$$
Figure~\no\ shows the long strip limiting behavior for the order
parameter.  The slopes are $\gamma / (2\nu_\perp)$ and $\gamma /
(2\nu_\parallel)$, respectively.  The large $L$ limit is easily
achieved, obtaining ${\gamma / (2\nu_\perp)} = 0.96 \pm 0.03$, in
accordance with theory.  The other limit is hard to reach, because of
the slow relaxation in transverse direction.  In any case, we found
${\gamma / (2\nu_\parallel)} = 0.37\pm 0.04$.

The order parameter or susceptibility at $T_c$ has an extra factor
which depends on the ratio $S$, e.g.,
$$ \Psi(T_c, L, M) = M^{-\beta/\nu_\perp}
\tilde \Psi(L^{\nu_\perp/\nu_\parallel}/M). \eqn $$
Moreover, the scaling function obeys $\tilde \Psi(S) \to
S^{-\nu_\parallel / (2\nu_\perp)}$ for $S \to \infty$, and $\tilde
\Psi(S) \to S^{1/2 - \beta/\nu_\perp}$ for $S \to 0$, as a consequence
of the limiting behaviors for very long strips [Eq.~(9) and (10)] and
the hyperscaling relation
$$2\beta + \gamma = \nu_\perp + \nu_\parallel. \eqn $$
The scaling form was tested for Ising model.$^{(12)}$  Previous
applications to anisotropic systems were not very
successful.$^{(8,14)}$  Figure~\no\ is a scaling plot with theoretical
values of exponents and $T_c = 1.41$.  Similar plots for $T=1.40$ and
$1.42$ show definite deviations from scaling.  This supports our
choice of $T_c$.  The asymptotic slopes for small and large scaling
variable $S = L^{1/3}/M$ are consistent with the expected value $-1/2$
and $-3/2$, respectively.

In Fig.~\no\ we plot similarly the susceptibility at $T_c=1.41$ in
scaling form.   The asymptotic scaling behavior is borne out,
$$ \chi(T_c, L, M) = M^{\gamma/\nu_\perp}
\tilde \chi(L^{\nu_\perp/\nu_\parallel}/M). \eqn $$
The scaling function $\tilde\chi(S)$ for large $S$ is
$\tilde\chi(S) \to const$, and for $S \to 0$ we have $\tilde\chi(S)
\to S^{\gamma/\nu_\perp}$.  The data are in full accord
with expectations.

Figure~\no\ is the scaling plot for the fourth-order cumulant,
$$ g(T_c, L, M) =
\tilde g (L^{\nu_\perp/\nu_\parallel}/M). \eqn $$
\noindent Large finite-size corrections are found here.  This is also
the case for the Ising model.$^{(12)}$

These scaling plots are good evidence that at least the exponent
ratios $\beta/\nu_\perp$ and $\nu_\parallel/\nu_\perp$ are in
agreement with the field-theoretic values.  Scaling off the critical
temperature will determine the exponents themselves.

\bigbreak
{\noindent\twelvebf 5. Anisotropic Scaling away from Critical Temperature}
\nobreak\smallskip\nobreak
We expect when $T \ne T_c$,
$$\Psi(T, L, M) = L^{-\beta/\nu_\parallel} \tilde
\Phi\bigl( L^{1/\nu_\parallel} (T-T_c)/T_c, S \bigr), \eqn $$
Leung$^{(9)}$ proposed a stronger scaling form when $S \to 0$.
Figure~\no\ is a scaling plot of the order parameter for $S =
2^{-8/3}$ with $T_c = 1.41$, assuming the field-theoretical exponents
$\beta = 1/2$, $\nu_\perp = 1/2$, and $\nu_\parallel = 3/2$.  The $T >
T_c$ branch (lower part) obeys scaling very well.  For the $T < T_c$
branch (upper part), deviation from scaling is large.  It appears that
the scaling region for $T < T_c$ is rather narrow.  The size effect is
further complicated by a $1/M$ correction due to the presence of
interfaces below $T_c$.  It is also not known how much of this can be
attributed to possible logarithmic corrections to scaling.  Comparing
with Leung's data,$^{(9,10)}$ we feel that his conclusion on data
collapse is somewhat too optimistic.

Our $1024\times 64$ order parameter data can be compatible with the
exponent $\beta = 1/2$, but only in a rather narrow critical region of
$\Delta T = 0.05$ (see Fig.~\no).  The data can be fitted to a power
$\beta \approx 0.3$ in a large temperature region ($\Delta T = 0.2$).
This is also roughly the value found in previous work on square
geometry.$^{(7,8)}$ However, the evidence for $\beta \approx 0.3$ is
not convincing since the extrapolated critical temperature would be
too low.

Figure~\no\ is a scaling plot for the susceptibility, $\chi'$, above
the critical temperature.  Good data collapse is obtained with the
theoretical exponents.  The asymptotic slope for large scaling
variable $M^2(T-T_c)/T_c$ is approximately $\gamma \approx 1$.  We
should not worry about the deviations from scaling for large values of
$T-T_c$.  Just like the order parameter, the data below the critical
temperature (not shown here) don't scale well.

\bigbreak
{\noindent\twelvebf 6. Conclusion}
\nobreak\smallskip\nobreak
Extensive computer simulation of the driven diffusive model is
performed.  The new data are consistent with the phenomenological
finite-size scaling theory with the set of exponents derived from
field-theoretic model.  Thus, this work supports the concept of
universality of the critical exponents.  The order parameter data
below the critical temperature are somewhat difficult to interpret.
The data at and above the critical temperature conform to standard
finite-size scaling.

\vfill\eject
{\noindent\twelvebf References}
\nobreak\smallskip\nobreak
{\frenchspacing

\item{1.} S. Katz, J. L. Lebowitz, and H. Spohn, {\it Phys. Rev. B\/}
{\bf 28}:1655 (1983).
\item{2.} S. Katz, J. L. Lebowitz, and H. Spohn, {\it J. Stat. Phys.}
{\bf 34}:497 (1984).

\item{3.} H. van Beijeren and L. S. Schulman, {\it Phys. Rev. Lett.}
{\bf 53}:806 (1984).

\item{4.} For reviews, see B. Schmittmann, {\it Int. J. Mod. Phys.}
{\bf 4}:2269 (1990); B. Schmittmann and R. K. P. Zia, in {\it Phase
Transitions and Critical Phenomena}, edited by C. Domb and
J. L. Lebowitz (Academic Press, New York, in press).

\item{5.} H. K. Janssen and B. Schmittmann, {\it Z. Phys. B\/}
{\bf 64}:503 (1986).
\item{6.} K.-t. Leung and J. Cardy, {\it J. Stat. Phys.}
{\bf 44}:567 (1986).

\item{7.} J. L. Vall\'es and J. Marro, {\it J. Stat. Phys.}
{\bf 49}:89 (1987).

\item{8.} A. Achahbar and J. Marro, {\it J. Stat. Phys.}, in press (1994).

\item{9.}  K.-t. Leung, {\it Phys. Rev. Lett.} {\bf 66}:453 (1991).

\item{10.}  K.-t. Leung, {\it Int. J. Mod. Phys. C\/} {\bf 3}:367 (1992).

\item{11.} N. Ito, {\it Int. J. Mod. Phys. C\/} {\bf 4}:525 (1993).

\item{12.} K. Binder and J.-S. Wang, {\it J. Stat. Phys.} {\bf 55}:87 (1989).

\item{13.} E. L. Praestgaard, H. Larsen, and R. K. P. Zia, {\it Europhys.
Lett.} {\bf 25}:447 (1994).

\item{14.} J.-S. Wang, K. Binder, and J. L. Lebowitz, {\it J. Stat. Phys.}
{\bf 56}:783 (1989).

}

\vfill\eject
{\noindent\twelvebf Figure Captions}
\nobreak\smallskip\nobreak
\medbreak
\parindent=13.6mm
\item{\bf Fig.~1.}  The locations of the susceptibility peaks plotted
against $L^{-2/3}$, for rectangular systems with $S=2^{-10/3}$
(diamonds, $M=32,\,64$), $S=1/8$ (squares, $M=16,\,32,\,64$),
$S=2^{-8/3}$ (circles, $M=16,\,32,\,64$), $S=1/4$ (triangles,
$M=16,\,32$), and for the square systems (pluses,
$M=32,\,64,\,128,\,256$).

\medbreak
\item{\bf Fig.~2.}  The fourth-order cumulant $g$ against scaling
variable $L^{2/3}(T-T_c)/T_c$, here $T_c=1.41$.  The system sizes are
$16\times 16$ (circles), $128\times 32$ (squares), and $1024 \times
64$ (triangles).

\medbreak
\item{\bf Fig.~3.}  Limiting behavior of the order parameter at $T_c$.
Circles are for $L\to\infty$ and squares are for $M\to \infty$.
The straight lines have slopes 1 and $1/3$, respectively.

\medbreak
\item{\bf Fig.~4.}  Scaling plot of the order parameter at the
critical point, $T_c = 1.41$.  Each set of data has a fixed $M$ value,
$M=4$ (solid circles), $M=8$ (open squares), $M=16$ (diamonds),
$M=32$ (pluses), $M=64$ (open circles), $M=128$ (squares),
$M=256$ (open triangles), $M=512$ (crosses), $M=1024$ (up triangles).

\medbreak
\item{\bf Fig.~5.}  Scaling plot of the susceptibility at the
critical point, $T_c=1.41$.  The symbols are the same as in Fig.~4.

\medbreak
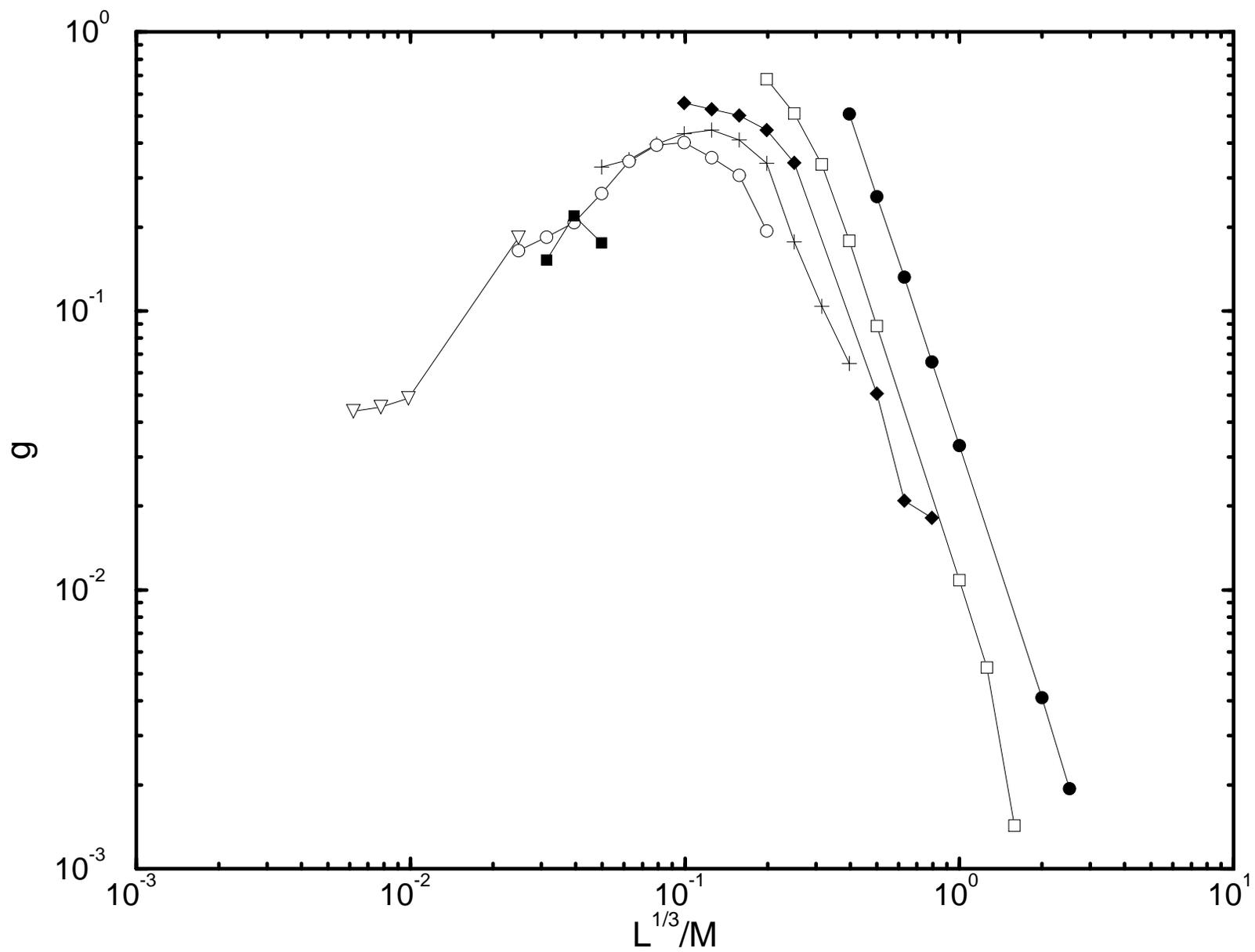
\item{\bf Fig.~6.}  Scaling plot of the fourth-order cumulant at the
critical point.  The symbols are the same as in Fig.~4.

\medbreak
\item{\bf Fig.~7.}  Scaling plot of the order parameter away from
critical point.    The system sizes are $16\times 16$
(circles), $128\times 32$ (open squares), and $1024 \times 64$
(triangles).

\medbreak
\item{\bf Fig.~8.} Order parameter to some power, $\Psi^2$ (circles)
and $\Psi^{10/3}$ (squares), versus temperature $T$.  The system size
is $1024\times 64$.

\medbreak
\item{\bf Fig.~9.}  Scaling plot of the susceptibility $\chi'$
above the critical temperature.  The system sizes are $8\times 16$
(circles), $64\times 32$ (open squares), and $512\times 64$
(triangles).

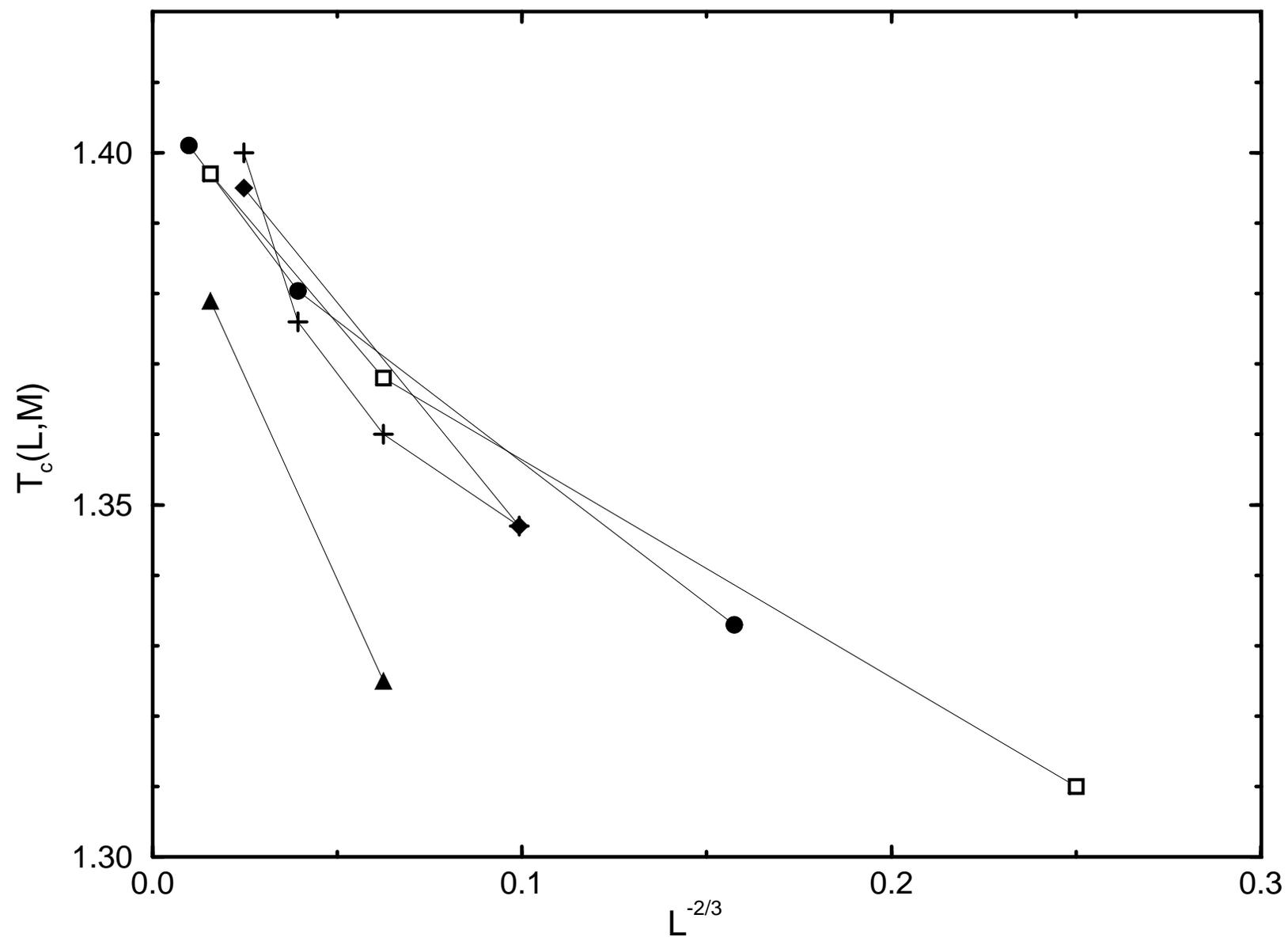
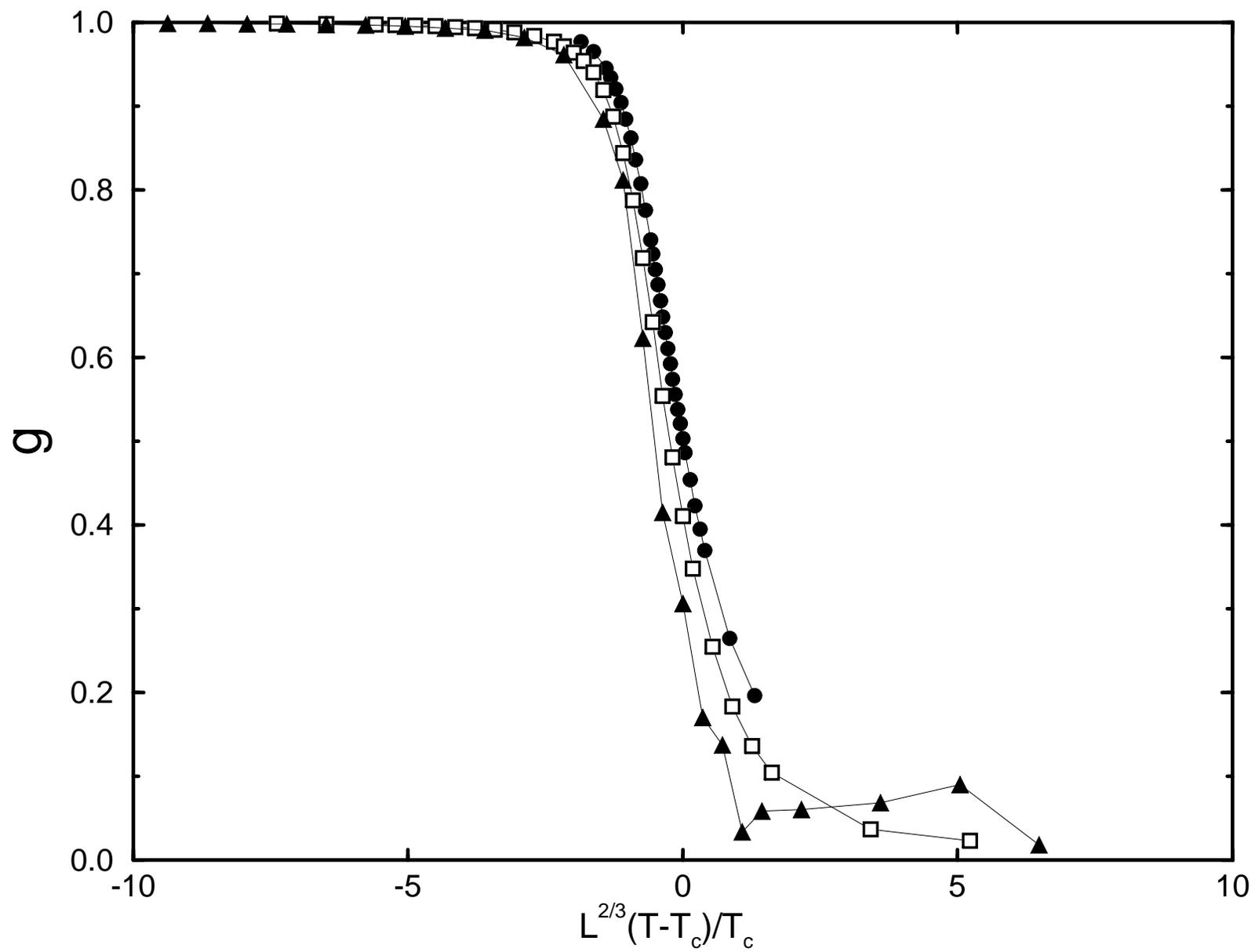
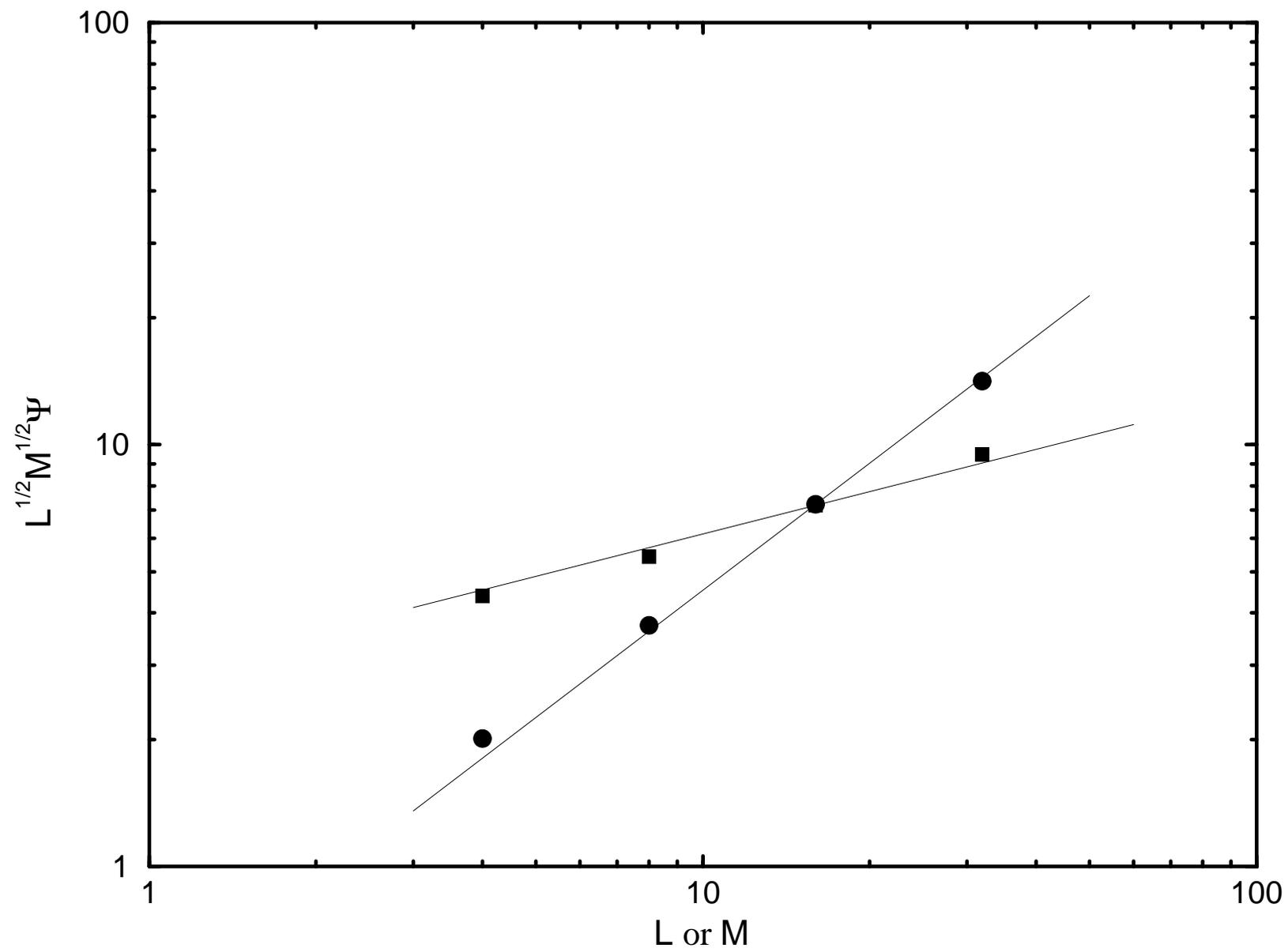
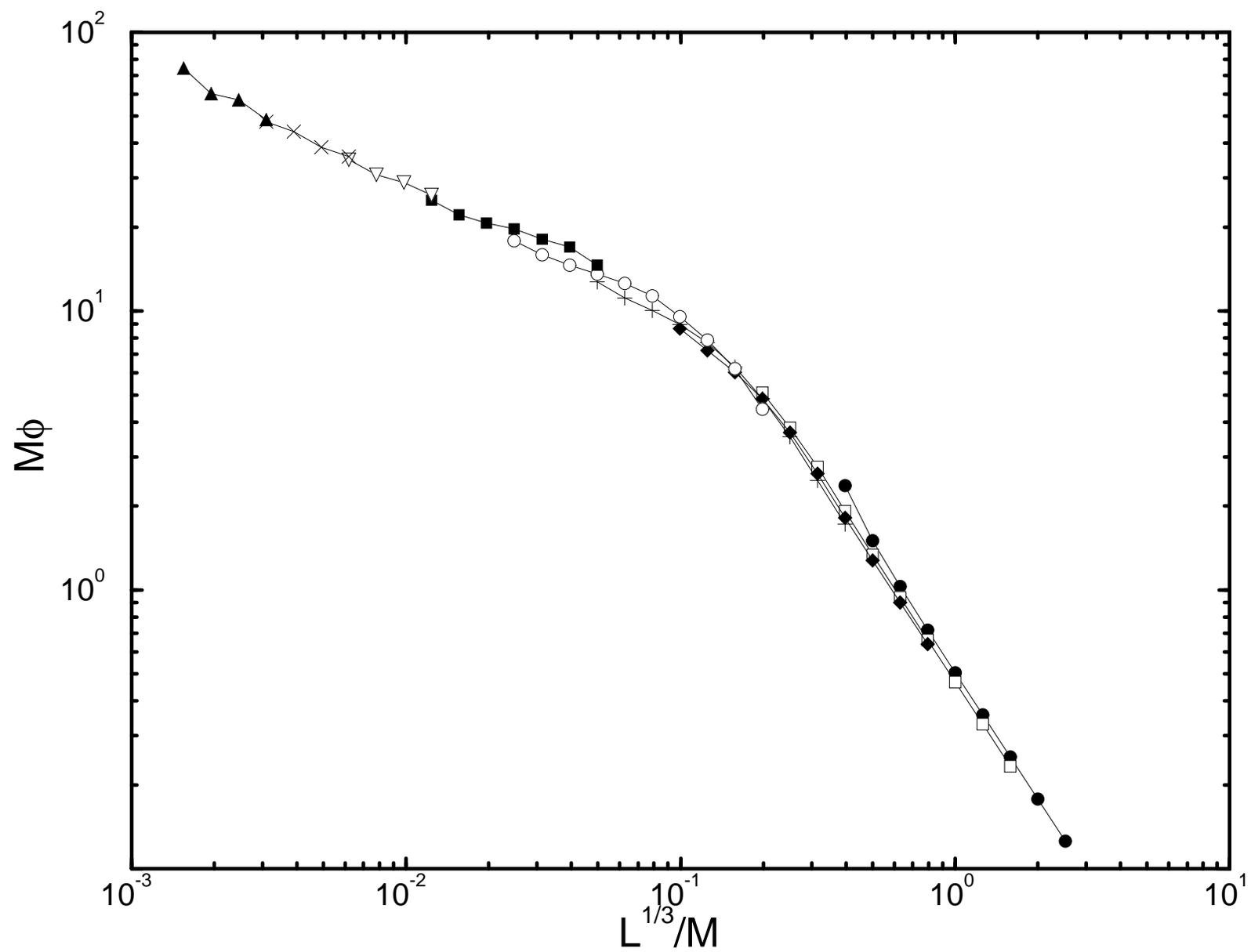
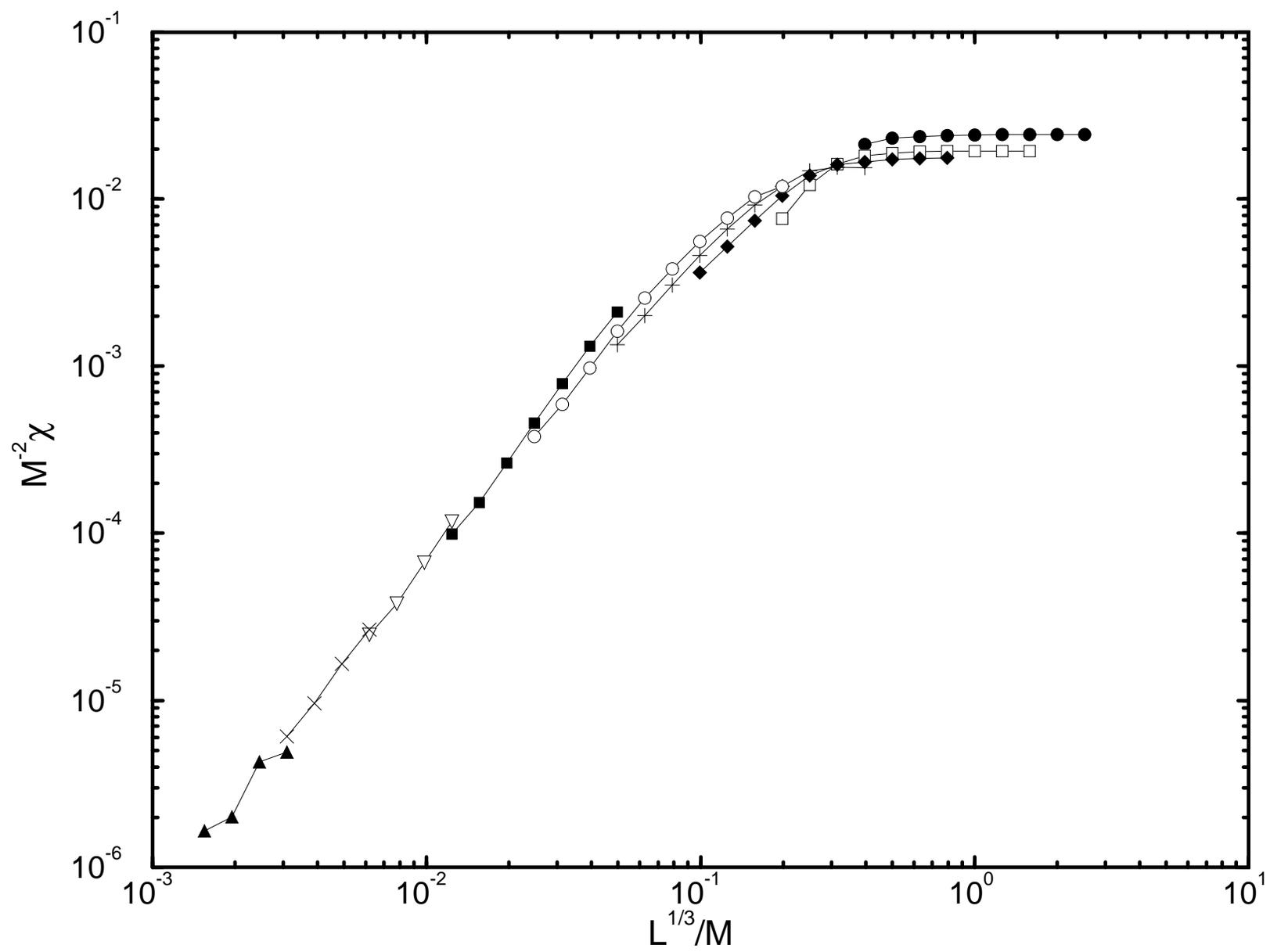
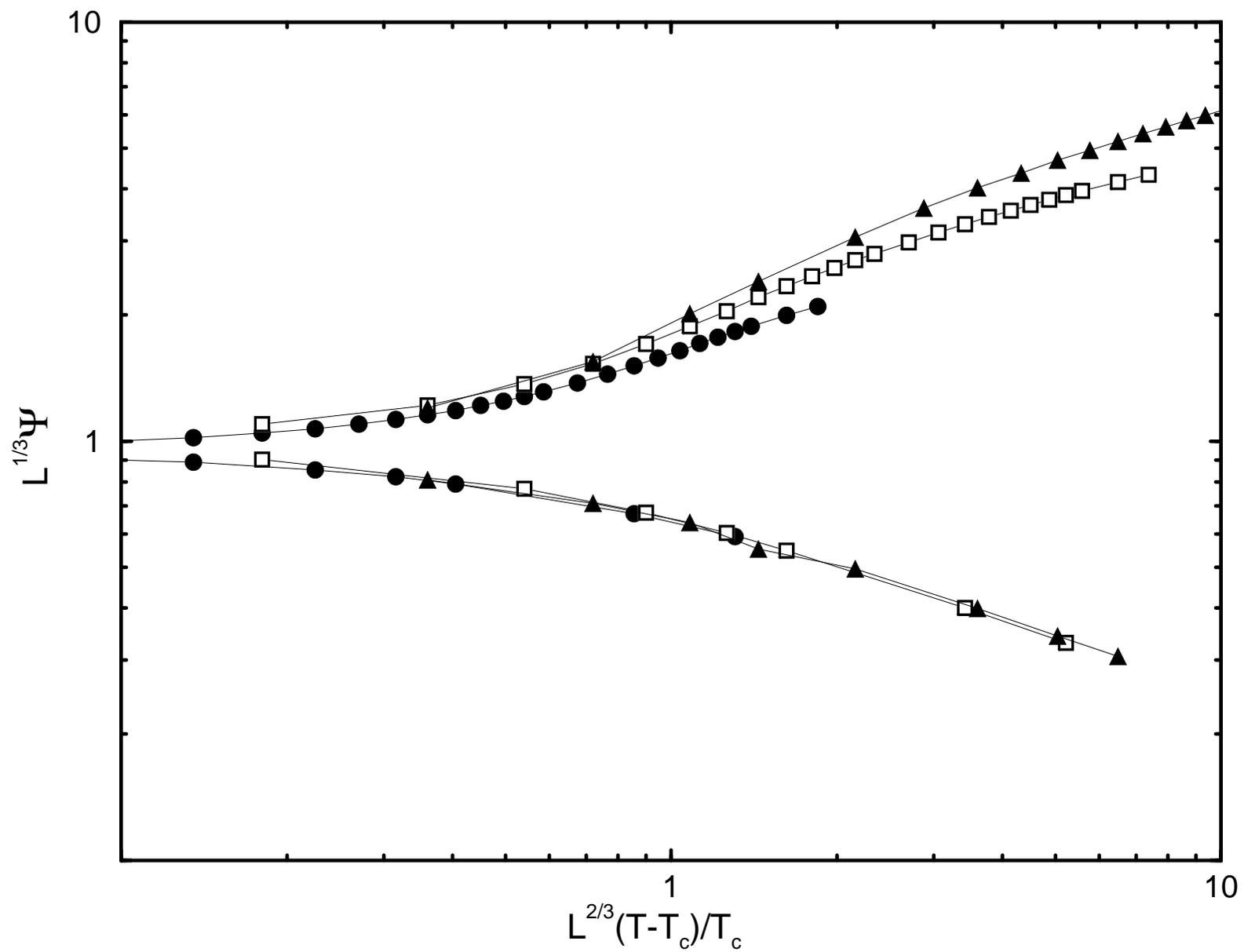
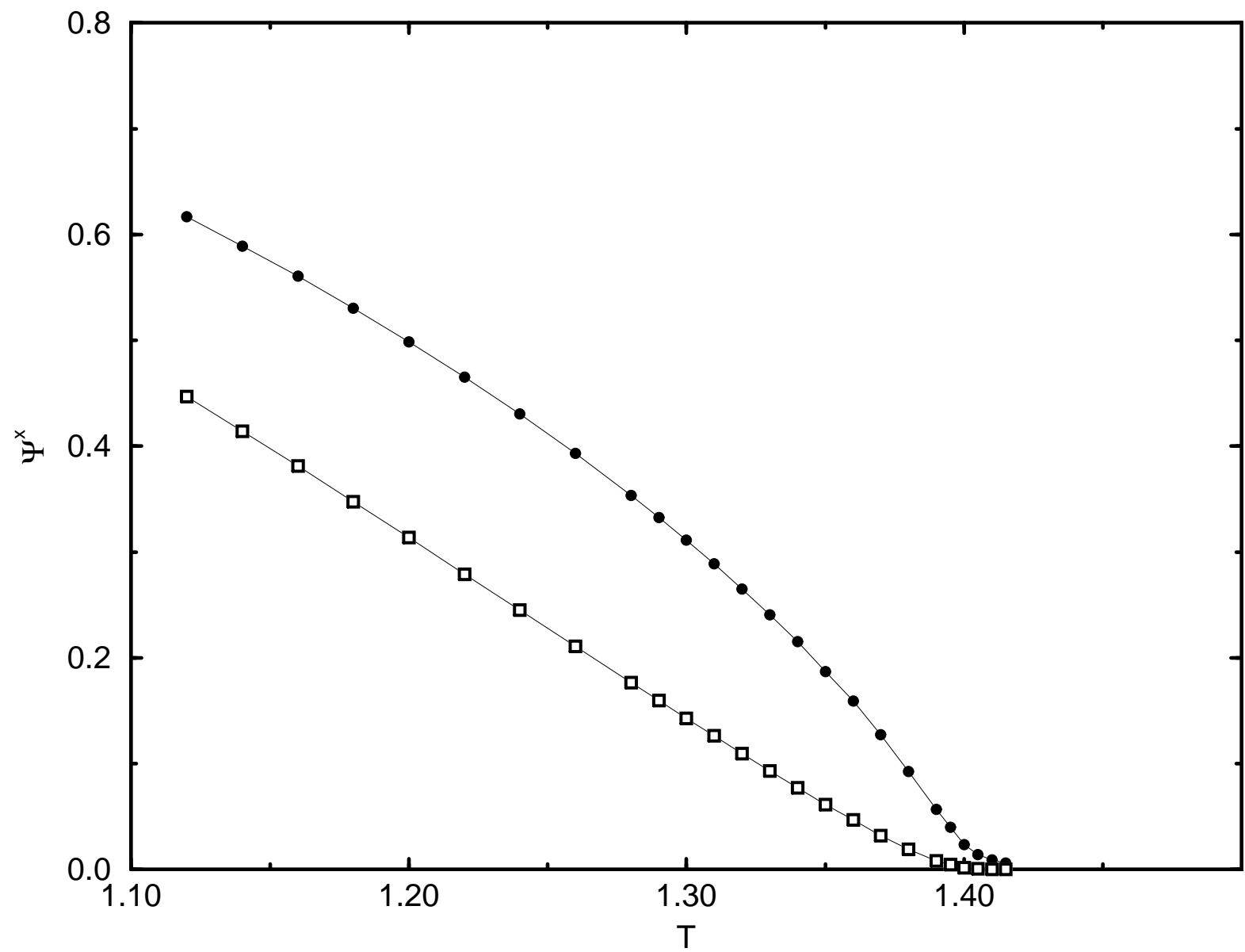
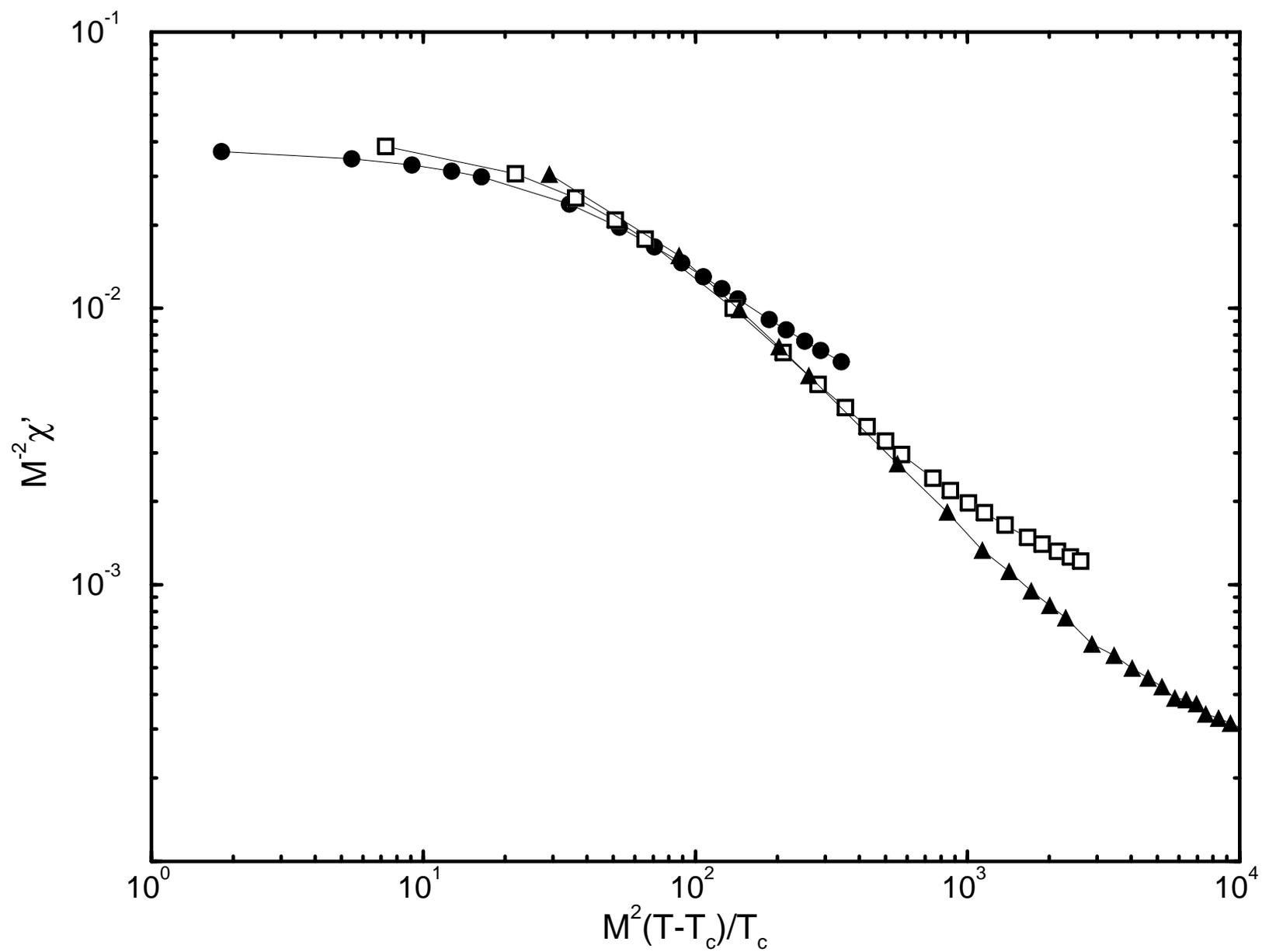
\bye